\documentclass[conference]{IEEEtran}
\usepackage{amssymb,amsfonts,amsmath,bbm}
\usepackage{dsfont}
\usepackage{flushend}
\usepackage{balance}
\usepackage{cite}
\usepackage{flushend}
\usepackage{mathtools}

\usepackage{multirow}
\usepackage{kantlipsum}
\usepackage{cuted}
\usepackage{tabu}
\usepackage{citesort}

\IEEEoverridecommandlockouts

\ifCLASSINFOpdf
   \usepackage[pdftex]{graphicx}
   \DeclareGraphicsExtensions{.pdf,.jpeg,.png}
\else
\fi
\usepackage{fixltx2e}

\usepackage{url}

\def\(({\left(}
\def\)){\right)}
\def\[[{\left[}
\def\]]{\right]}

\newcommand{\nn}{\nonumber \\}

\newcommand{\bT}{{\textbf {T}}}
\newcommand{\bO}{{\textbf {O}}}

\newcommand{\bu}{{\textbf {u}}}
\newcommand{\bV}{{\textbf {V}}}
\newcommand{\bF}{{\textbf {F}}}
\newcommand{\bY}{{\textbf {Y}}}

\newcommand{\bW}{{\textbf {W}}}
\newcommand{\bZ}{{\textbf {Z}}}

\newcommand{\bN}{{\textbf {N}}}

\newcommand{\bv}{{\textbf {v}}}

\newcommand{\bX}{{\textbf {X}}}
\newcommand{\bx}{{\textbf {x}}}

\newcommand{\by}{{\textbf {y}}}

\newcommand{\bR}{{\textbf {R}}}

\newtheorem{lemma}{\textbf{Lemma}}[section]
\newtheorem{proposition}{\textbf{Proposition}}[section]
\newtheorem{theorem}{\textbf{Theorem}}[section]

\newtheorem{remark}{\textbf{Remark}}[section]

\newcommand{\be}{\begin{equation}}
\newcommand{\ee}{\end{equation}}
\newcommand{\bea}{\begin{eqnarray}}
\newcommand{\eea}{\end{eqnarray}}

\newcommand{\EE}{{\mathbb{E}}}

\newcommand{\BEAS}{\begin{eqnarray*}}
\newcommand{\EEAS}{\end{eqnarray*}}
\newcommand{\BEA}{\begin{eqnarray}}
\newcommand{\EEA}{\end{eqnarray}}

\def\(({\left(}
\def\)){\right)}                       
\def\[[{\left[}
\def\]]{\right]}

\newcommand\smallO{
  \mathchoice
    {{\scriptstyle\mathcal{O}}}
    {{\scriptstyle\mathcal{O}}}
    {{\scriptscriptstyle\mathcal{O}}}
    {\scalebox{.7}{$\scriptscriptstyle\mathcal{O}$}}
  }

\newenvironment{talign*}
 {\csname align*\endcsname}
 {\endalign}
  
\newcommand{\simiid}[1][0pt]{%
  \mathrel{\raisebox{#1}{$\sim$}}%
}
\newcommand{\iid}{\overset{\text{\tiny i.i.d.}}{\simiid[-2pt]}}

\usepackage{color}

\usepackage{bm}
\graphicspath{{../figures/}}

\begin{document}
%
\title{The Mutual Information in Random Linear Estimation Beyond i.i.d. Matrices}

\author{
       \IEEEauthorblockN{Jean Barbier$^{\dagger*}$ and Nicolas Macris$^\dagger$}
        \IEEEauthorblockA{$\dagger$ Communication Theory Laboratory, EPFL, Switzerland.\\
$*$ International Center for Theoretical Physics, Trieste, Italy.}
\and
 \IEEEauthorblockN{Antoine Maillard  and Florent Krzakala}
        \IEEEauthorblockA{LPS ENS, CNRS, PSL, UPMC \& Sorbonne Universit\'e,\\ Paris, France.}
}



\maketitle

\begin{abstract}

  There has been definite progress recently in proving the
  variational single-letter formula given by the heuristic replica
  method for various estimation problems. In particular, the replica
  formula for the mutual information in the case of noisy linear
  estimation with random i.i.d. matrices, a problem with applications
  ranging from compressed sensing to statistics, has been proven
  rigorously. In this contribution we go beyond the restrictive
  i.i.d. matrix assumption and discuss the formula proposed by Takeda, Uda,
  Kabashima and later by Tulino, Verdu, Caire and Shamai who used the
  replica method. Using the recently introduced adaptive interpolation method and random matrix
  theory, we prove this formula for a relevant large sub-class 
  of rotationally invariant matrices.
\end{abstract}

\IEEEpeerreviewmaketitle
{\phantom{}}

Few problems are as ubiquitous in computer science as the one of
random linear estimation, that plays a fundamental role in machine
learning \cite{johnson1984extensions}, statistics
\cite{candes2006near} and communication \cite{2004random}. Computing
the information theoretic limitation for the estimation of a signal given
the knowledge of its random linear projections has many applications, e.g., compressed sensing
\cite{candes2006near}, code division multiple access
(CDMA)\cite{tanaka2002statistical} or error correcting codes
\cite{barron2010sparse,barbier2015approximate}. The problem is defined as follows:
Consider a {\it signal} vector $\bX\in \mathbb{R}^n$ with i.i.d. entries
distributed according to a ``prior'' $P_0$ over $\mathbb{R}$ with bounded support (an hypothesis that can be relaxed). One is given $m$ measurements
\begin{align}\label{eq:def_inference}
Y_\mu &= \sqrt{\frac{\lambda}{n}} (\bm{\Phi} \bX)_\mu +
Z_\mu \, ,\quad \mu = 1,\ldots,m\,,
\end{align}
in which $\lambda > 0$ is the {\it signal to noise ratio} (snr), $\bZ =
(Z_\mu)_{\mu=1}^m \iid \mathcal{N}(0,1)$ is a Gaussian noise
and $\bm{\Phi}\in\mathbb{R}^{m\times n}$ is the {\it measurement matrix}. We will
consider the ``high-dimensional limit'', namely $m,n \to \infty$ such that $\alpha \equiv m/n$ stays finite.

There has been a considerable amount of work on this model in the
case where $\bm{\Phi}$ is a random matrix whose elements
are i.i.d. standard Gaussian. In particular a pionnering work by Tanaka
\cite{tanaka2002statistical} using a statistical physics calculation
and the replica method \cite{mezard1990spin} proposed a
single-letter formula for the normalized mutual information 
between the measurements and the signal $i_n\equiv\frac{1}{n}I(\bX;\bY|\bm{\Phi})$.
%
%
The so-called Tanaka formula, orginally written for the CDMA problem
(where each element of $\bX$ is taken i.i.d. from $\pm 1$), has been
generalized and applied to many problems (e.g. in
\cite{krzakala2012statistical,tulino2013support}). After nearly $15$
years, it has been proven with different
approaches
\cite{BarbierDMK16,barbier_ieee_replicaCS,ReevesPfister2016,Barbier2017}, a
spectacular confirmation of the replica calculation.

In this paper, we endeavour to go beyond the very restrictive simple
i.i.d. measurement matrix assumption and consider instead a more complex situation where
$\bm{\Phi}$ is now taken from a non-trivial, and correlated, random
ensemble. This is both more relevant to practical cases, and more
realistic in terms of modeling real statistical situations.

\section{Result and related works}
The random linear estimation problem with correlated, non i.i.d. matrices has been considered, again with the replica method, and a
solution was proposed by Takeda, Uda and Kabashima in the case of
CDMA \cite{TakedaUdaKabashima2006} for matrices taken from a {\it rotationally invariant ensemble}. The replica formula was
later extended by Tulino, Caire, Verdu and Shamai
\cite{tulino2013support} for such similar ensembles in order to allow for more complicated priors
such as Gauss-Bernoulli ones. These works point to a generic conjecture, that we now describe, giving the
single-letter formula for the mutual information.  

Consider a measurement matrix
$\bm{\Phi} = \bO \bm{\Sigma} \bN^\intercal$
where $\bO$ and $\bN$ are both orthogonal matrices and $\bm\Sigma$ is diagonal (non-square if $\alpha \neq 1)$. The matrices $\bm\Sigma$, $\bN$, $\bO$ are independent
and $\bN$ is distributed uniformly,
according to the \emph{Haar measure} of its orthogonal
group, in which case $\bm{\Phi}$ is said to be \emph{right-rotationally invariant} (if $\bO$ is also Haar distributed the ensemble is simply called rotationally invariant). The matrix 
$\bR\equiv \frac{1}{n}\bm{\Phi}^\intercal \bm{\Phi} = \frac{1}{n}\bN \bm{\Sigma}^\intercal\bm{\Sigma}\bN^\intercal$ plays an important role. For general rotationally invariant ensembles its eigenvalues are not necesssarily i.i.d. but typically
are such that
$\frac{1}{n}\bm{\Sigma}^\intercal\bm{\Sigma}$ has a suitable limiting eigenvalue distribution (as in fact assumed in \cite{TakedaUdaKabashima2006}). 
The limit of the normalized mutual information is conjectured to be
\begin{align}\label{conjecture}
\lim_{n\to\infty} i_n &=\inf_{(E,r)\in\Gamma}i_{\rm RS}(E,r;\lambda)
\end{align}
where the so-called replica symmetric {\it potential} $i_{\rm RS}(E,r;\lambda)=i_{\rm RS}$ is defined as
\begin{align}
\textstyle i_{\rm RS} \equiv
I(X;\sqrt{r}X+Z)+\frac{1}{2}\int_0^{\lambda E} {\cal R}_\bR(-z)dz -
\frac{rE}2 \,.  \label{replica}
\end{align}
Here we have used the \emph{R-transform} of the matrix $\bR$ (see \cite{2004random} for an introduction to such transforms). $I(X;\sqrt{r}X+Z)$, that we simply denote $I(r)$ later on, is the mutual information for 
the scalar Gaussian channel $Y=\sqrt{r}X+Z$, $X\sim P_0$ and $Z\sim{\cal N}(0,1)$. Moreover, $\Gamma$ is the
set of critical points of the potential, or \emph{state evolution}
fixed points ($\rho \equiv \EE_{P_0}[X^2]$)
\begin{align*}
\Gamma  \equiv \big\{(E,r)\in[0,\rho]\times \mathbb{R}_+\,\big|\, E&= {\rm mmse}(X|\sqrt{r}\,X+Z), \nn \quad r&=\lambda {\cal R}_\bR(-\lambda E)\big\}\,.
\end{align*}

A virtue of the formula \eqref{replica} is that the details of the rotation invariant matrix ensemble only enter through the R-transform.

Note the slight difference with the potential written in \cite{tulino2013support}: In the potential \eqref{replica} a factor $1/2$, not present in \cite{tulino2013support}, multiplies both the integrated R-transform and $-r E$. This is because they consider the 
complex $\bm{\Phi}$ case while we consider the real case (nevertheless, we believe that our proof techniques could easily be generalized to include the complex case).

Interestingly, Manoel {\it et al} \cite{Manoeletal} and Reeves
\cite{Reeves} were able to formally re-derive equivalent results independently
for a sub-class of rotationally invariant matrices by considering multi-layered
estimation problems. This line of work combined with our present rigorous work is giving a lot of credibility to the replica conjecture for the general rotation invariant ensemble.

\subsection{Main result}
Our main result is a proof of the replica conjecture
(\ref{conjecture}) for a specific, but large, set of correlated
matrices. We hope it paves the way towards a completely general proof, as the non-rigourous replica calculation only 
assumes right-rotational invariance of $\bm{\Phi}$. We assume that the $m \times n $ matrix $\bm{\Phi}$ can be decomposed as follows: 
\begin{align}\label{eq:decomposition}
\bm{\Phi} = \bm{\Phi'} \bW
\end{align}
in which all elements of the $m \times n$ matrix $\bW$ are i.i.d. 
Gaussians ${\cal N}(0,1/n)$ with mean zero and variance $1/n$, $\bm{\Phi'}$ is a $m\times m$ random matrix, and $\bW$, $\bm{\Phi'}$ are independent. 
Concerning $\bm{\Phi'}$, our analysis is currently complete under the assumption that it is a product of a finite number of independent matrices, each with 
i.i.d. matrix-elements that are either bounded or standard Gaussians. 
The case of a product of i.i.d. standard Gaussian matrices constitutes an interesting example
that has been considered in \cite{Manoeletal}.

Our goal here is to give a rigorous proof of
 \eqref{conjecture} in the setting of matrices \eqref{eq:decomposition} with the independence assumptions for $\bm{\Phi'}$. 
 Some of our technical calculations are based on previous related works, and all of them will be discussed in a complete manner in a longer contribution.
\begin{theorem}[Replica formula]\label{thm:main_thm}
Assume that the prior $P_0$ has compact support or, in other words, the signal is bounded. Then with $n,m,\bm{\Phi},\bR,i_{\rm RS}$ defined as before (in particular $\bm{\Phi}$ satisfies \eqref{eq:decomposition} and the subsequent hypothesis), one has
\begin{align*}
\lim_{n\to\infty}i_n = {\adjustlimits \inf_{r\ge0}\sup_{E \in [0,\rho]}}i_{\rm RS}(E,r;\lambda)\,.
\end{align*}
\end{theorem}
\begin{remark}[Equivalent expressions of the replica formula]
The right hand side in the theorem above can also be written as $\inf_{(E,r)\in\Gamma}i_{\rm RS}(E,r;\lambda)$ (if the extremizers are not attained at the boundaries, which is the case when noise is present) or as $\inf_{E \in [0,\rho]}\sup_{r\ge0}i_{\rm RS}(E,r;\lambda)$. A proof of such equivalences, in the case of generalized linear estimation, is found in \cite{Barbier2017}.
\end{remark}

\begin{remark}[Right-rotation invariance of the ensemble]
Note that \eqref{eq:decomposition} implies the right-rotation invariance of $\bm{\Phi}$ because of the rotation invariance of $\bW$. Our result thus covers a sub-class of right-rotationally invariant matrices. 
\end{remark}

\begin{remark}[Relaxing the assumptions on ${\bm \Phi'}$]
The assumptions on $\bm{\Phi'}$ come from the fact that a complete 
rigorous analysis requires proving the concentration of the ``free energy'' of an {\it interpolating} model (see below; the free energy is equal to the mutual information up to a trivial additive constant). 
This involves the use of tools such as the McDiarmid bounded difference inequality and/or the Gaussian Poincar\'e inequality, which require some independence between degrees of freedom. 
If one {\it assumes} concentration of the free energy of the interpolating model then one can relax these assumptions to the following more general ones.  
It then suffices to assume that $\frac{1}{n}\bm{\Phi'}^\intercal \bm{\Phi'}$ has a well-defined, positively and compactly supported, asymptotic 
eigenvalue distribution in the limit $n \to \infty$. We also remark that concentration proofs of the free energy of the original and interpolated models are technically similar, however we have not established a purely logical implication between the two.
If such an implication holds one could also replace the independence assumption on ${\bm{\Phi'}}$ by an assumption of concentration of the free energy. 
\end{remark}

\begin{remark}[Relaxing the assumption on $P_0$]
Boundedness of the signal is again used to obtain concentration results for the free energy but it can presumably be removed 
using a limiting argument as in \cite{MiolaneXX}.
\end{remark}

\subsection{Related works}
There has been a lot of effort recently
\cite{BarbierDMK16,barbier_ieee_replicaCS,ReevesPfister2016,Barbier2017}
to prove the Tanaka formula for random
i.i.d. matrices. Our strategy in the present paper is to follow the {\it adaptive
interpolation method} introduced in
\cite{barbier_stoInt,2017arXiv170910368B}. This method, in particular,
has been used by the authors of \cite{Barbier2017} to reach a rigorous
demonstration of the replica formula for the mutual
information for the case of i.i.d. measurement matrix $\bm{\Phi}$,
in the more general situation of ``generalized linear
estimation'' i.e.,  with an {\it arbitrary} measurement channel (instead of just
a random additive noise as in (\ref{eq:def_inference})).  Some
steps of our current approach consequently follow similar ones in
\cite{Barbier2017} (but with key differences) and we will refer to
this work when necessary. We believe, in fact, that the approach presented
in the present paper could further be generalized as well to
generalized linear estimation with rotationally invariant matrices to
reach the formula conjectured by Kabashima in
\cite{Kabashima2007}. This is left for future work. 

Perhaps the most important consequence of the replica formula is that
it predicts the value of the minimum mean-square error (MMSE) in the
reconstruction of the signal $\bX$. In fact, it is conjectured (and
proved for Gaussian matrices \cite{barbier_ieee_replicaCS,ReevesPfister2016}) to be given by the value $E$ that
extremizes \eqref{replica}. While we consider here mainly the
information theoretic result, a large body of work has focused on
algorithmic approaches to random linear estimation, and investigated
whether the MMSE is efficiently (say, in polynomial time)
achievable. 

For Gaussian matrices, the most successful approach, so far, again
originated in statistical physics
\cite{thouless1977solution,mezard1989space} and is called approximate
message-passing (AMP) \cite{donoho2009message}. AMP is Bayes-optimal
and efficiently achieves the MMSE for a large set of
parameters, as proven in \cite{barbier_ieee_replicaCS}. There, however, might exist a region called ``hard''
where this is not the case, and polynomial algorithms improving on AMP
are not known. Whether there exists an efficient algorithm that is able to beat AMP in the hard region is widely considered to be a notoriously difficult problem (see e.g. \cite{REVIEWFLOANDLENKA} and reference therein).

For rotationally invariant matrices, different but related approaches
were proposed \cite{TakedaUdaKabashima2006,SAMP}. In particular, the
general expectation-propagation (EP) \cite{Minka} leads to a powerful
scheme in this context \cite{EP}. Recently Ma and Ping proposed a
variation of EP called OAMP \cite{ma2017orthogonal} specially adapted
to these matrices. Rangan, Schniter and Fletcher introduced a related
approach called VAMP \cite{VAMP} and showed that it follows the fixed
point equation (called state evolution) of the potential
\eqref{replica}. Interestingly, the multi-layer AMP algorithm of
Manoel {\it et al.} \cite{Manoeletal} also has the same fixed
point. 
Our result thus supports that OAMP, VAMP (and multi-layer AMP)
are Bayes-optimal and efficiently reach the MMSE in the ``easy''
region of random linear estimation with these correlated matrices,
just as AMP does in the case of i.i.d. Gaussian matrices.

\section{Proof by the adaptive interpolation method} 
We give here the main steps of the proof of Theorem~\ref{thm:main_thm}. We will use the adaptive interpolation method, introduced in \cite{barbier_stoInt}, and then applied in \cite{2017arXiv170910368B} and \cite{Barbier2017}. It is a powerful evolution of the interpolation method developed by Guerra and Toninelli in the context of spin glasses \cite{guerra2002thermodynamic}. Many steps of the proof follow the ones of \cite{Barbier2017}, and we will refer to them when necessary.




\subsection{Interpolating estimation problem}

Let us fix a sequence $s_n \in (0,1/2]$ that goes to $0$ as $n$ goes to infinity. Let $\epsilon = (\epsilon_1,\epsilon_2) \in [s_n,2 s_n]^2$ (so that $\epsilon$ actually depends on $n$, but we will drop this dependency for clarity). 

Let $E : [0,1] \to [0,\rho]$ and $r :  [0,1] \to\mathbb{R}_+$ be two continuous ``interpolation functions'' (that will later depend on $\epsilon$), and $R_1(t) \equiv \epsilon_1 + \int_0^t r(v) d v$, $R_2(t) \equiv \epsilon_2 + \int_0^t E(v) d v$ for $t \in [0,1]$. Consider the following two $t$-dependent observation channels for $i=1,\ldots,n$ and $\mu=1,\ldots,m$:

\begin{align}
	\begin{cases}
	Y_{t,\mu} &= \sqrt{\frac{\lambda(1-t)}{n}}(\bm{\Phi} \bX)_\mu+ \sqrt{\frac{\lambda}{n} R_2(t)} (\bm{\Phi'} \bV)_\mu  + Z_\mu\,,\\
	\widetilde Y_{t,i} &= \sqrt{R_1(t)} X_i + \widetilde Z_i\,. \label{channels}
	\end{cases}
\end{align} 
In the following we assume $\lambda = 1$, as it amounts to a scaling of $\bm{\Phi}$. Here $(\widetilde Z_i),(Z_\mu), (V_\mu)\iid \mathcal{N}(0,1)$ whereas $(X_i)\iid P_0$. The inference problem is to recover both $\bX=(X_i)_{i=1}^n$ and $\bV=(V_\mu)_{\mu=1}^m$ from the knowledge of the observations $\bY_t$, $\widetilde \bY_t$ and the matrix $\bm{\Phi}$ (and thus of $\bm{\Phi'}$ and $\bW$ too as the decomposition \eqref{eq:decomposition} is assumed to be known). 

In the Bayesian setting the posterior associated with this inference problem, written in the Gibbs-Boltzmann form, is
\begin{align}\label{post}
d P_{t,\epsilon}(\bx,\bv|\bY_t,\widetilde \bY_t,\bm{\Phi})=\frac{dP_0(\bx){\cal D}\bv e^{-{\cal H}(\epsilon,t,\bx,\bv;\bY_t,\widetilde \bY_t,\bm{\Phi})}}{\int dP_0(\bx){\cal D}\bv e^{-{\cal H}(\epsilon,t,\bx,\bv;\bY_t,\widetilde \bY_t,\bm{\Phi})}}\,,
\end{align}
where ${\cal D}\bv\equiv\prod_{\mu=1}^mdv_\mu (2\pi)^{-1/2}e^{-v_\mu^2/2}$ is the standard Gaussian measure, and we have defined the interpolating \emph{Hamiltonian} ${\cal H}={\cal H}(\epsilon,t,\bx,\bv;\bY_t,\widetilde \bY_t,\bm{\Phi})$ as
\begin{equation*}
\textstyle {\cal H}\equiv\frac12 \| \bY_t-\sqrt{\frac{1-t}{n}} \bm{\Phi}\bx\! -\! \sqrt{\frac{R_2(t)}{n}} \bm{\Phi'} \bv \|_2^2 + \frac12 \| \widetilde \bY_t-\sqrt{R_1(t)}\,\bx \|_2^2\,,
\end{equation*}

It is a simple exercise (see e.g. \cite{barbier_ieee_replicaCS}) to show that the normalized mutual information $i_n(t)\equiv\frac1nI(\bX,\bV;\bY_t,\widetilde \bY_t|\bm{\Phi})$ for the interpolation estimation problem is related to the posterior normalization (or partition function) through 
\begin{equation}\label{eq:def_mutual_info}
\textstyle i_{n,\epsilon}(t)= -\frac{1}{n}\EE\ln \int dP_0(\bx) {\cal D}\bv e^{-{\cal H}(\epsilon,t,\bx,\bv;\bY_t,\widetilde \bY_t,\bm{\Phi})} -\frac{\alpha+1}{2}\,.
\end{equation}
One can verify that this {\it interpolating mutual information} satisfises:
\begin{align}\label{i0i1}
\Big\{
\begin{array}{llll}
\!\!i_{n,\epsilon}(0)=i_n + \smallO_n(1),\\
\!\!i_{n,\epsilon}(1)=I(R_1(1))\!+\!\frac\alpha2 \EE_{X'} \!\ln (1 \!+\! R_2(1)X') \!+\! \smallO_n(1),
\end{array}		
\end{align}
where $X'\sim p'_n$, with $p'_n$ the empirical spectral distribution of the $m \times m$ matrix $\frac{1}{n}\bm{\Phi'}^\intercal \bm{\Phi'}$. Here, $\smallO_n(1) \to 0$ in the $n \to \infty$ limit, uniformly in $E,r,t,\epsilon$. The second term in the expression of $i_{n,\epsilon}(1)$ (sometimes refered to as a Shannon transform, see e.g. \cite{2004random}) is obtained using the celebrated ``log-det formula'' for the mutual information $\frac{1}{n}I(\bY_1;\bV|\bm{\Phi'})$ of an i.i.d. Gaussian input multiplied by the matrix $\bm{\Phi'}$ and under additive Gaussian noise, see e.g. \cite{2004random}.

Now a crucial step in our proof, that is a consequence of the particular form of the measurement matrix \eqref{eq:decomposition}, is that as $n$ grows, the second term in $i_{n,\epsilon}(1)$ can be replaced by an integrated $R$-transform. Denoting $G_{\bR}(x) \equiv \int_0^x \mathcal{R}_{\bR}(-u) d u$:
\begin{align}\label{13}
\textstyle i_{n,\epsilon}(1)=I(R_1(1))+\frac12 G_{\bR}(R_2(1)) + \smallO_n(1)\,,
\end{align}
where $\mathcal{R}_{\bR}(z)$ is the R-transform associated with the asymptotic spectrum of $\bR=\frac{1}{n}\bm{\Phi}^\intercal \bm{\Phi}$. We give the definition of this transform as well as the proof of \eqref{13} in the next section.

{\it A word about notations:} We define the Gibbs bracket $\langle-\rangle_{t,\epsilon}$ as the expectation w.r.t. the posterior \eqref{post}. In constrast, we denote by $\EE$ the joint expectation w.r.t. all \emph{quenched} variables (i.e. fixed by the realization of the problem), namely $(\bX,\bV,\bY_t,\widetilde \bY_t,\bm{\Phi'}, \bW)$, or equivalently w.r.t. $(\bX,\bV,\bZ,\widetilde \bZ,\bm{\Phi'}, \bW)$.

\subsection{Useful tools from random matrix theory}\label{sec:tools}
In this paragraph we show how to deduce \eqref{13} from \eqref{i0i1}. Note 
\begin{equation*}
\textstyle \mathbb{E}_{X'}\ln(1+R_2(1) X') = \mathbb{E}_{X'}\int_0^{R_2(1)} du \frac{X'}{1 + u X'}\,.
\end{equation*}
The result thus follows if the following relation is true :
\begin{equation}\label{R-S-relation}
\textstyle \mathcal{R}_{\bR} (- u) = \alpha \mathbb{E}_{X'}[\frac{X'}{1 + u X'}] +\smallO_n(1)\,.
\end{equation}
This is a well known relation in random matrix theory. For matrices of the form $\bR = \bW^\intercal \bT \bW$, where $\bW$ is a Gaussian $m\times n$ matrix with i.i.d. $\mathcal{N}(0, 1/n)$ elements, and $\bT=\frac{1}{n}\bm{\Phi'}^{\intercal}\bm{\Phi'}$ a non negative $m\times m$ matrix with a limiting spectral distribution, this was already shown by Marcenko and Pastur in 1967 \cite{marvcenko1967distribution} in the language of the Stieltjes transform. See also \cite{Silverstein-1995} for generalizations. Denoting by $g_\bR(z)$ and $g_\bT(z)$ ($z$ a complex number outside the 
specrum of the matrices)
the limiting Stieltjes transforms of the matrices $\bR$ and $\bT$, we have 
\cite{marvcenko1967distribution,Silverstein-1995} that the Marcenko-Pastur formula takes the form
\begin{align*}
\textstyle z (g_\bR(z))^2 + \alpha g_\bT( -1/g_\bR(z)) + (1-\alpha) g_\bR(z) = 0\,.
\end{align*}
Simple algebra then implies ($g_\bR^{-1}$ is the inverse function)
\begin{align*}
\textstyle g_\bR^{-1}(z) + z^{-1}  = - \alpha z^{-2} (g_\bT(-z^{-1}) - z)\,.
\end{align*}
Since by definition $\mathcal{R}_\bR(z) \equiv g_\bR^{-1}(-z) - z^{-1}$ and $g_\bT(z) \equiv \int  \frac{d\tau(x)}{x - z}$, $d\tau$ being the {\it limiting} 
eigenvalue distribution of $\bT$, this relation is nothing else than $\textstyle \mathcal{R}_\bR(-z) = \alpha \int  d\tau(x) \frac{x}{1 +zx}$
which is equivalent to \eqref{R-S-relation} when $n\to\infty$. We refer to the review \cite{speicher2009free} for a more modern discussion using free probability concepts.

\subsection{Mutual information variation}
In order to ``compare'' the potential \eqref{replica} with the mutual information, we use the trivial identity $$\textstyle i_n = i_{n,\epsilon}(0) + \smallO_n(1) = i_{n,\epsilon}(1) -\int_0^1 i_{n,\epsilon}'(t) 
dt + \smallO_n(1)$$ which becomes, using \eqref{13},
\begin{align}
	\textstyle i_n \!=\! I(R_1(1))+\frac12  G_{\bR}(R_2(1))-\int_0^1 i_{n,\epsilon}'(t) dt+\smallO_n(1)\,. \label{14}
\end{align}

We now evaluate $i_{n,\epsilon}'(t)$. Define $Q \equiv \frac{1}{n} \sum_{i=1}^n X_i x_i$, called the {\it overlap},
and the vector $\bu_{t}=(u_{t,\mu})_{\mu=1}^m$ with 
\begin{equation*}
\textstyle u_{t,\mu} \!\equiv\! \sqrt{\frac{1-t}{n}}(\bm{\Phi}(\bX\! - \!\bx))_\mu\!+\! \sqrt{\frac{R_2(t)}{n}} (\bm{\Phi'}(\bV - \bv) )_\mu \!+\! Z_\mu\,.
\end{equation*}
%
\begin{lemma}[Mutual information $t$-variation]\label{lemma:derivative}
For $t \in (0,1)$
\begin{align}
i_{n,\epsilon}'(t) = &~\frac{r(t)}{2} (\rho - \EE \langle Q \rangle_{t,\epsilon}) + \smallO_n(1) \nn &+ \frac{1}{2n^2} \EE  \big\langle \bZ^\intercal (\bm{\Phi'} \bm{\Phi'}^\intercal) \bu_{t} [E(t) - (\rho - Q) ] \big\rangle_{t,\epsilon}\label{i'}
\end{align}
\end{lemma}
The proof of this lemma is very similar to the one found in \cite{Barbier2017}. The idea is to write explicitly the derivative $i_{n,\epsilon}'(t)$ and then to integrate by parts the quenched Gaussian variables $\bV$ and $\bW$, before using the {\it Nishimori identity}. This identity is a consequence of Bayes rule and the fact that we consider the {\it optimal Bayesian setting}, namely that all hyperparameters in the problem such as the snr and $P_0$ are known and used when defining the posterior, see \cite{barbier_stoInt,Barbier2017,REVIEWFLOANDLENKA}.

\subsection{Overlap concentration}

The next lemma essentially states that the overlap concentrates around its mean, and plays a key role in our proof. 
The proof technique for Bayesian 
inference has been developped in \cite{Macris2007,korada2010,korada2009exact,barbier_ieee_replicaCS} and is akin to the analysis reviewed for example in 
\cite{talagrand2010meanfield1}. The point however here is that in Bayesian inference overlap concentration can be proved in the whole phase diagram.
We will refer to \cite{barbier_stoInt,Barbier2017} where the analysis has been made quite generic. We now write explicitely the dependency of $R_1(t,\epsilon)$ and $R_2(t,\epsilon)$ on $\epsilon$.

\begin{lemma}[Overlap concentration] \label{concentration}
Assume that for any $t \in (0,1)$ the map $\epsilon=(\epsilon_1,\epsilon_2) \in [s_n,2s_n]^2 \mapsto R(t,\epsilon)=(R_1(t,\epsilon),R_2(t,\epsilon))$ is a $\mathcal{C}^1$ diffeomorphism with Jacobian determinant greater or equal to $1$. Then one can find a sequence $s_n$ going to $0$ slowly enough such that there exist positive constants $C$ and $\gamma$ that only depend on the support and moments of $P_0$ and on $\alpha$, and such that:
 \begin{align*}
\textstyle \frac{1}{s_n^2}\int_{[s_n,2s_n]^2} d\epsilon\int_0^1 dt\, \mathbb{E}\big\langle \big(Q - \mathbb{E}\langle Q\rangle_{t, \epsilon}\big)^2 \big\rangle_{t, \epsilon}  \leq C n^{-\gamma}.
\end{align*}
\end{lemma}

We refer to \cite{barbier_stoInt,Barbier2017} for a detailed proof (in the case where $\bm{\Phi'}$ is the identity matrix). For the present model under the assumptions on $\bm{\Phi'}$, this follows from Gaussian Poincar\'e and McDiarmid inequalities much as in \cite{Barbier2017}. As a consequence of this result, together with Lemma.~\ref{lemma:derivative}, we obtain (using continuity and boundedness properties of the functions $I$ and $G_{\bR}$, see again \cite{Barbier2017} for more details):
\begin{lemma}[Fundamental identity]\label{lemma:fund_identity}
Assume $\epsilon\mapsto R(t,\epsilon)$ satisfies the hypotheses of Lemma~\ref{concentration}, and choose $s_n\to 0$ according to this lemma. Assume that for all $t \in [0,1]$ and $\epsilon \in [s_n,2s_n]^2$ we have $E(t,\epsilon) = \rho - \mathbb{E}\langle Q \rangle_{t,\epsilon}$. Then: 
\begin{eqnarray*}
\textstyle i_n = \frac{1}{s_n^{2}} \int_{[s_n,2s_n]^2} d\epsilon \big\{I(\int_0^1 r(t,\epsilon)  dt) + \frac{1}{2} G_{\bR}(\int_0^1 E(t,\epsilon)  dt) 
\\ \textstyle   - \frac{1}{2} \int_0^1E(t,\epsilon) r(t,\epsilon)  dt \big\} + \smallO_n(1)\,,
\end{eqnarray*}
in which $\smallO_n(1)$ is uniform in the choice of the functions $E,r$.
\end{lemma}

\subsection{Upper and lower bounds}

Similar bounds can be found in \cite{Barbier2017,2017arXiv170910368B}. We will often refer to \cite{Barbier2017} for more details. We first prove the upper bound:
\begin{proposition}$\limsup_{n \to \infty} i_n \leq {\adjustlimits \inf_{r \geq 0} \sup_{E \in [0,\rho]}} i_{\rm RS}(E,r;1)$.
\end{proposition}
\begin{IEEEproof}
Choose first $r(t) = r \geq 0$ a fixed value. We then fix $R = (R_1,R_2)$ as the solution $R(t,\epsilon)=(\epsilon_1+rt,\epsilon_2+\int_0^t E(s,\epsilon)ds)$ to the first order differential equation: $\partial_t R_1(t) = F_1$, $\partial_t R_2(t) = F_2(t,R(t))$, and $R(0) = \epsilon$, with $F_1=r$, $F_2(t,R(t))=\rho - \mathbb{E}\langle Q \rangle_{t,\epsilon}$ (it is easy to show that $F_2$ is in $[0,\rho]$, and thus $E$ too). One can check (see \cite{Barbier2017}) that this ODE satisfies the hypotheses of the Cauchy-Lipschitz theorem. As $F=(F_1,F_2)$ is continuous and admits continuous partial derivatives, $R(t,\epsilon)$ is ${\cal C}^1$ (in both arguments). By the Liouville formula, the Jacobian determinant $J_{n,\epsilon}(t)$ of $\epsilon \mapsto R(t,
\epsilon)$ satisfies $J_{n,\epsilon}(t) = \exp \{\int_0^t\partial_{R_2} F_2(s,R(s,\epsilon))ds\} \geq 1$. Indeed, the partial derivative 
$\partial_{R_2} F_2$ is non-negative, see Prop.~6 of \cite{Barbier2017}. Also, as this Jacobian never cancels, and as $\epsilon \mapsto R(t,
\epsilon)$ is injective (by unicity of $R(t,
\epsilon)$), it is a diffeomorphism by the inversion theorem. Recall $G_{\bR}(x) \equiv \int_0^x \mathcal{R}_{\bR}(-u) d u$ and \eqref{replica}. Then Lemma.~\ref{lemma:fund_identity} implies: $$\textstyle i_n = \frac{1}{s_n^{2}}  \int_{[s_n,2s_n]^2} d\epsilon \, i_{\rm RS}(\int_0^1 E(t,\epsilon)dt,r;1)+ \smallO_n(1)\,,$$ that directly gives the desired bound.
\end{IEEEproof}

We now turn to the lower bound:
\begin{proposition}
$\liminf_{n \to \infty} i_n \geq {\adjustlimits \inf_{r \geq 0} \sup_{E \in [0,\rho]}} i_{\rm RS}(E,r;1)$.
\end{proposition}
\begin{IEEEproof}
Fix $R$ as the solution $R(t,\epsilon)=(\epsilon_1+\int_0^t r(s,\epsilon)ds,\epsilon_2+\int_0^t E(s,\epsilon)ds)$ to the following Cauchy problem: $\partial_t R_1(t) = F_1(t,R(t))=\mathcal{R}_\bR(\mathbb{E}\langle Q\rangle_{t,\epsilon}-\rho)$, $\partial_t R_2(t) = F_2(t,R(t)) = \rho-\mathbb{E}\langle Q\rangle_{t,\epsilon}$ and $R(0) = \epsilon$. Let us denote this equation $\partial_t R(t) = F(t,R(t))$ ($F$ also depends on $n$). Note that this implies that the solutions verify $E(t,\epsilon)=\rho-\mathbb{E}\langle Q\rangle_{t,\epsilon} \in [0,\rho]$ and $r(t,\epsilon) \geq 0$. It is possible to verify (see the details in a similar case in \cite{Barbier2017}) that $F(t,R)$ is a bounded $\mathcal{C}^1$ function of $R$, and thus the Cauchy-Lipschitz theorem implies that $R(t,\epsilon)$ is a $\mathcal{C}^1$ function of both $t$ and $\epsilon$. The Liouville formula for the Jacobian determinant $J_{n,\epsilon}(t)$ of the map $\epsilon \mapsto R(t,\epsilon)$ yields $J_{n,\epsilon}(t) = \exp\{\int_0^t\partial_{R_1} F_1(s,R(s,\epsilon))ds+\int_0^t\partial_{R_2} F_2(s,R(s,\epsilon))ds\} \geq 1$. 
Indeed, one can show (see again \cite{Barbier2017}) that both partial derivatives (in the exponential) are non-negative for all $s \in (0,1)$. By the same arguments as in the previous bound, for any $t$, the map $\epsilon \mapsto R(t,\epsilon)$ a ${\cal C}^1$ diffeomorphism. All hypotheses of Lemma.~\ref{lemma:fund_identity} are verified. It leads to
\begin{eqnarray*}
\textstyle i_n = \frac{1}{s_n^{2}} \int_{[s_n,2s_n]^2} d\epsilon \big\{I(\int_0^1 r(t,\epsilon)  dt) + \frac{1}{2} G_{\bR}(\int_0^1 E(t,\epsilon)  dt) 
\\ \textstyle    - \frac{1}{2} \int_0^1E(t,\epsilon) r(t,\epsilon)  dt \big\} + \smallO_n(1)\,.
\end{eqnarray*}
$I$ is a concave function (see \cite{Barbier2017}), and so is $x \mapsto G_{\bR}(x)$. Indeed, by identity \eqref{R-S-relation}, we have $G_{\bR}''(x)\leq 0$. Jensen's inequality thus yields (and recalling \eqref{replica})
\begin{eqnarray*}
\textstyle i_n \geq \frac{1}{s_n^{2}} \int d\epsilon \int_0^1  dt \, i_{\rm RS}(E(t,\epsilon),r(t,\epsilon);1) + \smallO_n(1)\,. 
\end{eqnarray*}
Notice $i_{\rm RS}(E(t,\epsilon),r(t,\epsilon);1)=\sup_{E\in[0,\rho]}i_{\rm RS}(E,r(t,\epsilon);1)$. Indeed, $g_r : E \mapsto i_{\rm RS}(E,r;1)$ is also concave (by concavity of $G_{\bR}$), with derivative $g_r'(E) = \frac{1}{2}\mathcal{R}_\bR(-E)-\frac{r}{2}$. By definition of the solution $R(t,\epsilon)$, $g_{r(t,\epsilon)}'(E(t,\epsilon)) = 0$ for any $(t,\epsilon)$, so by concavity $g_{r(t,\epsilon)}$ reaches its maximum at $E(t,\epsilon)$. Thus we finally obtain
\begin{talign*}
i_n &\geq \frac{1}{s_n^{2}} \int d\epsilon \int_0^1  dt \, \sup_{E \in [0,\rho]} i_{\rm RS}(E,r(t,\epsilon);1) + \smallO_n(1) \nn
& \geq  \inf_{r \geq 0} \sup_{E \in [0,\rho]} i_{\rm RS}(E,r;1)+ \smallO_n(1)\,.
\end{talign*}
Taking the $\liminf$, it ends the proof of Theorem \ref{thm:main_thm}.
\end{IEEEproof} 

%
%

\section*{Acknowledgment}
We acknowledge funding from the ERC under the European Union's FP7
Grant Agreement 307087-SPARCS, the SNSF grant 200021-156672, and the ANR PAIL. We also thank Olivier Lev\^eque and Sundeep Rangan for helpful discussions.

\bibliographystyle{IEEEtran}
\bibliography{refs}

\begin{thebibliography}{10}
\providecommand{\url}[1]{#1}
\csname url@samestyle\endcsname
\providecommand{\newblock}{\relax}
\providecommand{\bibinfo}[2]{#2}
\providecommand{\BIBentrySTDinterwordspacing}{\spaceskip=0pt\relax}
\providecommand{\BIBentryALTinterwordstretchfactor}{4}
\providecommand{\BIBentryALTinterwordspacing}{\spaceskip=\fontdimen2\font plus
\BIBentryALTinterwordstretchfactor\fontdimen3\font minus
  \fontdimen4\font\relax}
\providecommand{\BIBforeignlanguage}[2]{{%
\expandafter\ifx\csname l@#1\endcsname\relax
\typeout{** WARNING: IEEEtran.bst: No hyphenation pattern has been}%
\typeout{** loaded for the language `#1'. Using the pattern for}%
\typeout{** the default language instead.}%
\else
\language=\csname l@#1\endcsname
\fi
#2}}
\providecommand{\BIBdecl}{\relax}
\BIBdecl

\bibitem{johnson1984extensions}
W.~B. Johnson and J.~Lindenstrauss, ``Extensions of lipschitz mappings into a
  hilbert space,'' \emph{Contemporary mathematics}, 1984.

\bibitem{candes2006near}
E.~J. Candes and T.~Tao, ``Near-optimal signal recovery from random
  projections: Universal encoding strategies?'' \emph{IEEE Transactions on
  Information Theory}, vol.~52, no.~12, pp. 5406--5425, Dec 2006.

\bibitem{2004random}
A.~M. Tulino and S.~Verd{\'u}, \emph{Random matrix theory and wireless
  communications}.\hskip 1em plus 0.5em minus 0.4em\relax Now Publishers Inc,
  2004, vol.~1.

\bibitem{tanaka2002statistical}
T.~Tanaka, ``A statistical-mechanics approach to large-system analysis of cdma
  multiuser detectors,'' \emph{IEEE Transactions on Information Theory},
  vol.~48, no.~11, pp. 2888--2910, Nov 2002.

\bibitem{barron2010sparse}
A.~R. Barron and A.~Joseph, ``Toward fast reliable communication at rates near
  capacity with gaussian noise,'' in \emph{2010 IEEE International Symposium on
  Information Theory}, June 2010, pp. 315--319.

\bibitem{barbier2015approximate}
J.~Barbier and F.~Krzakala, ``Approximate message-passing decoder and capacity
  achieving sparse superposition codes,'' \emph{IEEE Transactions on
  Information Theory}, vol.~63, no.~8, pp. 4894--4927, Aug 2017.

\bibitem{mezard1990spin}
M.~M{\'e}zard, G.~Parisi, and M.-A. Virasoro, \emph{Spin glass theory and
  beyond.}\hskip 1em plus 0.5em minus 0.4em\relax World Scientific Publishing
  Co., Inc., Pergamon Press, 1987.

\bibitem{krzakala2012statistical}
F.~Krzakala, M.~M{\'e}zard, F.~Sausset, Y.~Sun, and L.~Zdeborov{\'a},
  ``Statistical-physics-based reconstruction in compressed sensing,''
  \emph{Phys. Rev. X}, vol.~2, p. 021005(18), May 2012.

\bibitem{tulino2013support}
A.~M. Tulino, G.~Caire, S.~Verd{\'u}, and S.~Shamai, ``Support recovery with
  sparsely sampled free random matrices,'' \emph{IEEE Transactions on
  Information Theory}, vol.~59, no.~7, pp. 4243--4271, July 2013.

\bibitem{BarbierDMK16}
J.~Barbier, M.~Dia, N.~Macris, and F.~Krzakala, ``The mutual information in
  random linear estimation,'' in \emph{2016 54th Annual Allerton Conference on
  Communication, Control, and Computing}, 2016.

\bibitem{barbier_ieee_replicaCS}
J.~Barbier, N.~Macris, M.~Dia, and F.~Krzakala, ``Mutual information and
  optimality of approximate message-passing in random linear estimation,''
  \emph{arXiv:1701.05823}, 2017.

\bibitem{ReevesPfister2016}
G.~Reeves and H.~D. Pfister, ``The replica-symmetric prediction for compressed
  sensing with gaussian matrices is exact,'' vol. arxiv:1607.02524.

\bibitem{Barbier2017}
\BIBentryALTinterwordspacing
J.~Barbier, F.~Krzakala, N.~Macris, L.~Miolane, and L.~Zdeborov{\'a}, ``Optimal
  errors and phase transitions in high-dimensional generalized linear models,''
  in \emph{Proceedings of the 31st Conference On Learning Theory}, ser.
  Proceedings of Machine Learning Research, vol.~75.\hskip 1em plus 0.5em minus
  0.4em\relax PMLR, July 2018, pp. 728--731. [Online]. Available:
  \url{http://arxiv.org/abs/1708.03395}
\BIBentrySTDinterwordspacing

\bibitem{TakedaUdaKabashima2006}
K.~Takeda, S.~Uda, and Y.~Kabashima, ``Analysis of cdma systems that are
  characterized by eigenvalue spectrum,'' \emph{EPL (Europhysics Letters)},
  vol.~76, no.~6, p. 1193, 2006.

\bibitem{Manoeletal}
A.~Manoel, F.~Krzakala, M.~M\'ezard, and L.~Zdeborov\'a, ``Multi-layer
  generalized linear estimation,'' in \emph{2017 IEEE International Symposium
  on Information Theory (ISIT)}, 2017, pp. 2098--2102.

\bibitem{Reeves}
G.~Reeves, ``Additivity of information in multilayer networks via additive
  gaussian noise transforms,'' vol. abs/1710.04580, 2017.

\bibitem{MiolaneXX}
M.~{Lelarge} and L.~{Miolane}, ``{Fundamental limits of symmetric low-rank
  matrix estimation},'' \emph{ArXiv e-prints}, Nov. 2016.

\bibitem{barbier_stoInt}
\BIBentryALTinterwordspacing
J.~Barbier and N.~Macris, ``The adaptive interpolation method: a simple scheme
  to prove replica formulas in bayesian inference,'' \emph{Probability Theory
  and Related Fields}, Oct 2018. [Online]. Available:
  \url{https://doi.org/10.1007/s00440-018-0879-0}
\BIBentrySTDinterwordspacing

\bibitem{2017arXiv170910368B}
J.~{Barbier}, N.~{Macris}, and L.~{Miolane}, ``{The Layered Structure of Tensor
  Estimation and its Mutual Information},'' in \emph{47th Annual Allerton
  Conference on Communication, Control, and Computing}, 2017.

\bibitem{Kabashima2007}
Y.~Kabashima, ``Inference from correlated patterns: a unified theory for
  perceptron learning and linear vector channels,'' \emph{Journal of Physics:
  Conference Series}, vol.~95, no.~1, p. 012001, 2008.

\bibitem{thouless1977solution}
D.~J. Thouless, P.~W. Anderson, and R.~G. Palmer, ``Solution of`solvable model
  of a spin glass','' \emph{Philosophical Magazine}, vol.~35, no.~3, p.
  593–601, 1977.

\bibitem{mezard1989space}
M.~M{\'e}zard, ``The space of interactions in neural networks: Gardner's
  computation with the cavity method,'' \emph{Journal of Physics A:
  Mathematical and General}, vol.~22, no.~12, pp. 2181--2190, 1989.

\bibitem{donoho2009message}
D.~L. Donoho, A.~Maleki, and A.~Montanari, ``Message-passing algorithms for
  compressed sensing,'' \emph{Proceedings of the National Academy of Sciences},
  vol. 106, no.~45, pp. 18\,914--18\,919, Nov 2009.

\bibitem{REVIEWFLOANDLENKA}
L.~Zdeborov\'a and F.~Krzakala, ``Statistical physics of inference: thresholds
  and algorithms,'' \emph{Advances in Physics}, vol.~65, no.~5, p. 453, 2016.

\bibitem{SAMP}
B.~{\c{C}}akmak, O.~Winther, and B.~H. Fleury, ``{S-AMP:} approximate message
  passing for general matrix ensembles,'' vol. arxiv:1405.2767.

\bibitem{Minka}
T.~P. Minka, ``Expectation propagation for approximate bayesian inference,'' in
  \emph{Proceedings of the Seventeenth Conference on Uncertainty in Artificial
  Intelligence}, ser. UAI'01, 2001, pp. 362--369.

\bibitem{EP}
M.~Opper and O.~Winther, ``Expectation consistent approximate inference,''
  \emph{Journal of Machine Learning Research}, vol.~6, p. 2177–2204, 2005.

\bibitem{ma2017orthogonal}
J.~Ma and L.~Ping, ``Orthogonal amp,'' \emph{IEEE Access}, vol.~5, pp.
  2020--2033, 2017.

\bibitem{VAMP}
S.~Rangan, P.~Schniter, and A.~K. Fletcher, ``Vector approximate message
  passing,'' vol. arxiv:1610.03082.

\bibitem{guerra2002thermodynamic}
F.~Guerra and F.~L. Toninelli, ``The thermodynamic limit in mean field spin
  glass models,'' \emph{Communications in Mathematical Physics}, vol. 230,
  no.~1, pp. 71--79, 2002.

\bibitem{marvcenko1967distribution}
V.~A. Mar{\v{c}}enko and L.~A. Pastur, ``Distribution of eigenvalues for some
  sets of random matrices,'' \emph{Mathematics of the USSR-Sbornik}, vol.~1,
  no.~4, p. 457, 1967.

\bibitem{Silverstein-1995}
J.~W. Silverstein, ``Strong convergence of the empirical distribution of
  eigenvalues of large dimensional random matrices,'' \emph{Journal of
  Multivariate Analysis}, vol.~55, no.~2, pp. 331--339, 1995.

\bibitem{speicher2009free}
R.~Speicher, ``Free probability theory,'' \emph{arXiv preprint
  arXiv:0911.0087}, 2009.

\bibitem{Macris2007}
N.~Macris, ``Griffith-{K}elly-{S}herman correlation inequalities: A useful tool
  in the theory of error correcting codes,'' \emph{IEEE Transactions on
  Information Theory}, vol.~53, no.~2, pp. 664--683, Feb 2007.

\bibitem{korada2010}
S.~B. Korada and N.~Macris, ``Tight bounds on the capacity of binary input
  random cdma systems,'' \emph{IEEE Transactions on Information Theory},
  vol.~56, no.~11, pp. 5590--5613, Nov 2010.

\bibitem{korada2009exact}
------, ``Exact solution of the gauge symmetric p-spin glass model on a
  complete graph,'' \emph{Journal of Statistical Physics}, vol. 136, no.~2, pp.
  205--230, 2009.

\bibitem{talagrand2010meanfield1}
M.~Talagrand, \emph{Mean field models for spin glasses: Volume I: Basic
  examples}.\hskip 1em plus 0.5em minus 0.4em\relax Springer Science \&
  Business Media, 2010, vol.~54.

\end{thebibliography}
\flushend

 \newpage
 \onecolumn
 \appendix
 \subsection{Nishimori identity}\label{appendix:nishimori}

\begin{lemma}\label{prop:nishimori}
	Let $(X,Y)$ be a couple of random variables on a polish space $E$. For a given $k \in \mathbb{N}^*$, let $(X^{(i)})_{i=1}^k$ be i.i.d random variables from the distribution (conditional over $Y$) $P(X = \cdot|Y)$. Denote $\langle - \rangle$ the expectation with respect to this probability distribution, and $\EE$ the expectation with respect to the probability measure of $(X,Y)$. Then, for all $f : E^{k+1} \to \mathbb{R}$ continuous and bounded
	\begin{align*}
	\EE \langle f(Y,X^{(1)},\ldots,X^{(k)})  \rangle = \EE \langle f(Y,X^{(1)},\ldots,X^{(k-1)},X)  \rangle\,.
	\end{align*}
\end{lemma}

\begin{IEEEproof}
This is a trivial consequence of Bayes formula: 
\begin{align*}
\EE_{X,Y} \langle f(Y,X^{(1)},\cdots,X^{(k-1)},X) \rangle = \EE_{Y} \EE_{X|Y} \langle f(Y,X^{(1)},\cdots,X^{(k-1)},X)= \EE \langle f(Y,X^{(1)},\cdots,X^{(k)})\rangle\,.
\end{align*}
\end{IEEEproof}

\subsection{Proof of Lemma~\ref{lemma:derivative}}\label{appendix:proof_derivative}

The proof is done in two steps. First, we show the following formula:
\begin{align}\label{eq:derivative_first_step}
&i_{n,\epsilon}'(t) = \frac{r(t)}{2} (\rho - \EE \langle Q \rangle_{t,\epsilon}) + \frac{1}{2n^2}  \sum_{\mu=1}^m \sum_{\nu=1}^m  \EE  \Big\langle Z_\mu (\bm{\Phi'} \bm{\Phi'}^\intercal)_{\mu \nu} u_{t,\nu} \Big[E(t) - \Big(\frac{1}{n}\sum_{i=1}^n X_i^2 - Q \Big) \Big] \Big\rangle_{t,\epsilon} \,.
\end{align}
We will then conclude using the concentration of $\frac{1}{n} \sum_{i=1}^n X_i^2$ on $\rho$ by the central limit theorem as $n \to \infty$. 

Recall that we defined the Gibbs bracket 
\begin{align}
\langle A(\bx,\bv)\rangle_{t,\epsilon} = \frac{\int dP_0(\bx) {\cal D}\bv e^{-{\cal H}(\epsilon,t,\bx,\bv;\bY_t,\widetilde \bY_t,\bm{\Phi})} A(\bx,\bv)}{\int dP_0(\bx) {\cal D}\bv e^{-{\cal H}(\epsilon,t,\bx,\bv;\bY_t,\widetilde \bY_t,\bm{\Phi})}}\,.
\end{align}
From this and the definition of $i_{n,\epsilon}(t)$ (\ref{eq:def_mutual_info}), one gets
\begin{align}\label{eq:in_derivative}
i_{n,\epsilon}'(t) = \frac{1}{n}\EE\big[ {\cal H}'(\epsilon,t,\bX,\bV;\bY_t,\widetilde \bY_t,\bm{\Phi})\ln {\cal Z}\big]+
\frac{1}{n}\mathbb{E}\big\langle {\cal H}'(\epsilon,t,\bx,\bv;\bY_t,\widetilde \bY_t,\bm{\Phi})\big\rangle_{t,\epsilon}\,,
\end{align}
where the partition function $\cal Z$ and Hamiltonian derivative with respect to $t$ read
\begin{align}
{\cal H}'(\epsilon,t,\bX,\bV;\bY_t,\widetilde \bY_t,\bm{\Phi})
&=\frac{1}{2}\sum_{\mu=1}^m Z_\mu\Big(\sqrt{\frac{1}{n(1-t)}}(\bm{\Phi}\bX)_\mu - \frac{E(t)}{\sqrt{nR_2(t)}} (\bm{\Phi'} \bV)_\mu\Big)-\frac12 \sum_{i=1}^n\widetilde Z_i \frac{r(t)}{\sqrt{R_1(t)}} X_i \,,\\
{\cal Z}={\cal Z}(\epsilon,t,\bY_t,\widetilde \bY_t,\bm{\Phi})&\equiv\int dP_0(\bx) {\cal D}\bv e^{-{\cal H}(\epsilon,t,\bx,\bv;\bY_t,\widetilde \bY_t,\bm{\Phi})}\,.
\end{align}

The Nishimori identity (Lemma~\ref{prop:nishimori}) directly implies  
\begin{align}
\mathbb{E}\big\langle {\cal H}'(\epsilon,t,\bx,\bv;\bY_t,\widetilde \bY_t,\bm{\Phi})\big\rangle_t = \mathbb{E} {\cal H}'(\epsilon,t,\bX,\bV;\bY_t,\widetilde \bY_t,\bm{\Phi}) = 0\,.
\end{align}

We now compute $\EE[ \widetilde Z_iX_i\ln {\cal Z}]$. Using a Gaussian integration by parts, which reads for any real function $f$ with continuous derivative $\EE[\widetilde Z_if(\widetilde Z_i)]=\EE[f'(\widetilde Z_i)]$ for $\widetilde Z_i \sim {\cal N}(0,1)$, we obtain the first term of \eqref{eq:in_derivative} as
	\begin{align}		
		&-\frac1{2n}\frac{r(t)}{\sqrt{R_1(t)}}\sum_{i=1}^n\EE \Big[X_i \widetilde Z_i \ln \int dP_0(\bx) {\cal D}\bv e^{-{\cal H}(\epsilon,t,\bx,\bv;\bY_t,\widetilde \bY_t,\bF)}  \Big]\nonumber\\
		=~&-\frac1{2n}\frac{r(t)}{\sqrt{R_1(t)}}\sum_{i=1}^n\EE \Big[X_i \widetilde Z_i \ln \int dP_0(\bx) {\cal D}\bv\, 
		e^{
			{\text{term}}_1(\bv,\bx)- \frac{1}{2} \sum_i\big(\sqrt{R_1(t)}(X_{i}-x_i) + \widetilde Z_i \big)^2
		}  
	\Big]\nonumber \\
		=~& \frac1{2n}\frac{r(t)}{\sqrt{R_1(t)}}\sum_{i=1}^n\EE \big[X_i \big\langle \sqrt{R_1(t)}(X_i - x_i) +\widetilde Z_i \big\rangle_{t,\epsilon} \big]
	\nn
	=~& \frac{r(t)\rho}{2}-\frac{r(t)}{2}
		\EE 
		\Big\langle
			\frac1n\sum_{i=1}^{n} X_i x_i
		\Big\rangle_{t,\epsilon} \nonumber\\
		=~ &\frac{r(t)}{2}(\rho-
		\EE \langle Q \rangle_{t,\epsilon})\,.
	\end{align}
%
In the same way, an integration by parts with respect to $V_i \sim \mathcal{N}(0,1)$ yields 
\begin{align}\label{eq:derivative_term_2}
	&-\frac1{2n}\frac{E(t)}{\sqrt{nR_2(t)}}\sum_{\mu=1}^m\EE\Big[ Z_\mu (\bm{\Phi'} \bV)_\mu \ln {\cal Z}\Big]\nonumber\\ 
=~ &-\frac1{2n}\frac{E(t)}{\sqrt{nR_2(t)}}\sum_{\mu=1}^m \sum_{i=1}^n\EE\Big[ Z_\mu \Phi'_{\mu i} V_i  \ln\int dP_0(\bx){\cal D}\bv e^{-\frac12 \sum_{\nu} \big(\sqrt{\frac{1-t}{n}} (\bm{\Phi}(\bX-\bx))_\nu + \sqrt{\frac{R_2(t)}{n}} (\bm{\Phi}'(\bV-\bv))_\nu+Z_\nu \big)^2+{\text{term}}_2(\bx)}\Big] \nonumber \\
	=~ &\frac{E(t)}{2n^2} \sum_{\mu=1}^m \sum_{\nu=1}^m \sum_{i=1}^n\EE\Big[  Z_\mu \Phi'_{\mu i} \Phi'_{\nu i}\Big\langle\sqrt{\frac{1-t}{n}} (\bm{\Phi}(\bX-\bx))_\nu + \sqrt{\frac{R_2(t)}{n}} (\bm{\Phi'}(\bV-\bv))_\nu+Z_\nu  \Big\rangle_{t,\epsilon}\Big] \nonumber \\
	 =~ &\frac{E(t)}{2n^2}  \sum_{\mu=1}^m \sum_{\nu=1}^m\EE\Big[   Z_\mu (\bm{\Phi'} \bm{\Phi'}^\intercal)_{\mu \nu}\langle u_{t,\nu} \rangle_{t,\epsilon}\Big]\,.
\end{align}
Let us now look at the final term we need to compute. By our hypothesis \eqref{eq:decomposition}, this term reads, using again a Gaussian integration by part but this time with respect to $W_{ji}\sim{\cal N}(0,1/n)$, 
\begin{align*}
&\frac{1}{2n}\sqrt{\frac{1}{n(1-t)}} \sum_{\mu=1}^m\EE\Big[Z_\mu(\bm{\Phi'} \bW \bX)_\mu \ln{\cal Z}\Big] \nonumber\\
= ~ & \frac1{2n} \sqrt{\frac{1}{n(1-t)}}\sum_{\mu=1}^m\sum_{i,j=1}^n\EE\Big[Z_\mu \Phi'_{\mu j}W_{ji} X_i \ln\int dP_0(\bx){\cal D}\bv e^{-\frac12 \sum_{\nu} \big(\sqrt{\frac{1-t}{n}} (\bm{\Phi'\bW}(\bX-\bx))_\nu + \sqrt{\frac{R_2(t)}{n}} (\bm{\Phi}'(\bV-\bv))_\nu+Z_\nu \big)^2+{\text{term}}_2(\bx)}\Big]\nonumber\\
= ~ & -\frac1{2n^3}\sum_{\mu,\nu=1}^m\sum_{i,j=1}^n\EE\Big[Z_\mu \Phi'_{\mu j}\Phi'_{\nu j} X_i \Big\langle(X_i-x_i)\Big(\sqrt{\frac{1-t}{n}} (\bm{\Phi}(\bX-\bx))_\nu + \sqrt{\frac{R_2(t)}{n}} (\bm{\Phi'}(\bV-\bv))_\nu+Z_\nu\Big)  \Big\rangle_{t,\epsilon}\Big]\nonumber\\
= ~ & -\frac1{2n^2}\sum_{\mu,\nu=1}^m\EE\Big[Z_\mu  (\bm{\Phi'} \bm{\Phi'}^\intercal)_{\mu \nu} \Big\langle u_{t,\nu}\Big(\frac1n\sum_{i=1}^n X_i^2 - \frac1n\sum_{i=1}^n X_ix_i \Big) \Big\rangle_{t,\epsilon}\Big]\,.
\end{align*}
Combining all three terms leads to \eqref{eq:derivative_first_step}.

We now go to the last step. By adding and substracting a term to \eqref{eq:derivative_first_step} we reach
\begin{align}\label{eq:derivative_second_step}
i_{n,\epsilon}'(t) = &\frac{r(t)}{2} (\rho - \EE \langle Q \rangle_{t,\epsilon}) + \frac{1}{2n^2}  \sum_{\mu=1}^m \sum_{\nu=1}^m  \EE  \big\langle Z_\mu (\bm{\Phi'} \bm{\Phi'}^\intercal)_{\mu \nu} u_{t,\nu} \big[E(t) - (\rho - Q ) \big] \big\rangle_{t,\epsilon} \nn
&+\frac{1}{2n^2}  \sum_{\mu=1}^m \sum_{\nu=1}^m  \EE  \Big\langle Z_\mu (\bm{\Phi'} \bm{\Phi'}^\intercal)_{\mu \nu} u_{t,\nu} \Big(\rho - \frac{1}{n}\sum_{i=1}^n X_i^2 \Big) \Big\rangle_{t,\epsilon} \,.
\end{align}
Using the Cauchy-Schwarz inequality we obtain that the last term can be bounded as
\begin{align}
\frac1{n^2}\Big| \EE  \Big\langle \bZ^{\intercal} (\bm{\Phi'} \bm{\Phi'}^\intercal) \bu_{t} \Big(\rho - \frac{1}{n}\sum_{i=1}^n X_i^2 \Big) \Big\rangle_{t,\epsilon}\Big| \le \Big\{ \frac1{n^4}\EE  \Big\langle \Big(\bZ^{\intercal} (\bm{\Phi'} \bm{\Phi'}^\intercal) \bu_{t}\Big)^2\Big\rangle_{t,\epsilon} \EE\Big[\Big(\rho - \frac{1}{n}\sum_{i=1}^n X_i^2 \Big)^2\Big] \Big\}^{1/2}\,.\label{33}
\end{align}
As the $X_i$ are independent the central limit theorem implies that $\EE[(\rho - \frac{1}{n}\sum_{i=1}^n X_i^2 )^2]={\cal O}(1/n)$. Thus it remains to show that the multiplicative term in front is bounded:
\begin{align}
&\frac1{n^4}\EE  \big\langle\big( \bZ^{\intercal} (\bm{\Phi'} \bm{\Phi'}^\intercal) \bu_{t}\big)^2\big\rangle_{t,\epsilon} \le \frac{1}{n^4}\EE\big\langle \|\bZ\|^2 \|\bu_t\|^2 \|\bm{\Phi'} \bm{\Phi'}^\intercal\|_{\rm F}^2\rangle_{t,\epsilon} \nn
\le~& \frac{1}{n^4}\sqrt{\EE\big\langle \|\bZ\|^4 \|\bu_t\|^4\big\rangle_{t,\epsilon} \EE\big[ \|\bm{\Phi'} \bm{\Phi'}^\intercal\|_{\rm F}^4\big]}
\le\frac{1}{n^4}\sqrt{\sqrt{\EE\big[ \|\bZ\|^8\big]\EE\big\langle \|\bu_t\|^8\big\rangle_{t,\epsilon}}  \EE\big[ \|\bm{\Phi'} \bm{\Phi'}^\intercal\|_{\rm F}^4\big]}={\cal O}(1)\,. \label{34}
\end{align}
The last equality follows from the following observations. By construction of $\bm{\Phi'}$, $\EE[ \|\bm{\Phi'} \bm{\Phi'}^\intercal\|_{\rm F}^4]^{1/2} = {\cal O}(n^2)$. Moreover, as $\bZ$ is a $m$-dimensional Gaussian vector with i.i.d. components $\EE[ \|\bZ\|^8]^{1/4}={\cal O}(n)$. Finally, the Nishimori identity leads to $\EE[\langle\|\bu_t\|^8\rangle_{t,\epsilon}]^{1/4}={\cal O}(n)$. This claim is proven using a consequence of the triangle inequality:
\begin{align*}
\forall \bx,\by \in \mathbb{R}^n \qquad \|x+y\|^8 \leq 2^7 (\|x^8\| + \|y^8\|)\,,
\end{align*}
which is combined with the Nishimori identity: 
\begin{align}
\EE[\langle\|\bu_t\|^8\rangle_{t,\epsilon}] =&~	\EE\Big\langle\Big\|\sqrt{\frac{1-t}{n}} \bm{\Phi}(\bX-\bx) + \sqrt{\frac{R_2(t)}{n}} \bm{\Phi'}(\bV-\bv)+\bZ\Big\|^8\Big\rangle_{t,\epsilon} \nn
\le&~ 2^7 \EE[ \|\bZ\|^8] + 2^{22}(1-t)^4\EE[ \|\frac{1}{\sqrt{n}}\bm{\Phi}\bX \|^8] + 2^{22}R_2(t)^4\EE[\| \frac{1}{\sqrt{n}}\bm{\Phi'}\bV\|^8]\,.\label{35}
\end{align}

One can now use that both $\frac{1}{n} \bm{\Phi}^\intercal\bm{\Phi}$ and $\frac{1}{n} \bm{\Phi'}^\intercal\bm{\Phi'}$ have almost surely bounded Euclidian (or Frobenius) norm when $n \to \infty$. This implies that there exists $C > 0$ such that
\begin{align*}
\EE[ \|\frac{1}{\sqrt{n}}\bm{\Phi}\bX \|^8] \leq C \EE[ \|\bX \|^8] \,.
\end{align*}
Moreover $\EE[ \|\bX \|^8] = {\cal O}(n^4)$ because we assumed the prior distribution $P_0$ to be compactly supported. The same argument can be conducted for bounding $\EE[ \|\frac{1}{\sqrt{n}}\bm{\Phi'}\bV \|^8]$ since $\EE[ \|\bV \|^8] = {\cal O}(n^4)$ as $\bV$
 is a standard Gaussian vector. 

\end{document}